\documentclass[traditabstract]{aa} 
\usepackage{longtable}
\usepackage{latexsym}
\usepackage{amssymb}
\usepackage{lscape}
\usepackage[]{natbib}

\usepackage{color}

\usepackage{graphicx}
\usepackage{txfonts}
\def\lsim{\mathrel{\rlap{\lower 3pt \hbox{$\sim$}} \raise 2.0pt \hbox{$<$}}}
\def\gsim{\mathrel{\rlap{\lower 3pt \hbox{$\sim$}} \raise 2.0pt \hbox{$>$}}}

   \title{Automated bar detection in local disc galaxies from the SDSS}
   \subtitle{The colors of bars}
   \author{Guido Consolandi \inst{1}
   }
   \authorrunning{G. Consolandi}
   \institute{
    Dipartimento di Fisica G. Occhialini, Universit\`a di Milano-Bicocca, Piazza della Scienza 3, I-20126 Milano, Italy\\
    \email{guido.consolandi@mib.infn.it} 
}
\begin{document}
\date{}

  \abstract{This paper describes an automatic isophotal fitting procedure that succeeds, 
  without the support of any visual inspection of neither the images nor the ellipticity/position-angle radial profiles, at extracting a 
  fairly pure sample of barred late-type galaxies (LTGs) among thousands of optical images from the Sloan Digital Sky Survey (SDSS). The procedure relies on the methods described in Consolandi et al. (2016) to robustly extract the photometrical
  properties of a large sample of local SDSS galaxies and is tailored to extract bars on the basis of their well-known peculiarities in their P.A. and ellipticity profiles.
It has been run on a sample of 5853 galaxies in the Coma and Local supercluster. The procedure extracted for each galaxy a color, an ellipticity 
and a position angle radial profile of the ellipses fitted to the isophotes. Examining automatically the profiles of 922 face-on late-type galaxies (B/A$>0.7$) the procedure 
found that $\sim 36 \%$ are barred. The local bar fraction strongly increases with stellar mass. The sample of barred galaxies is used to 
construct a set of template radial color profiles in order to test the impact of the barred galaxy population on the average color profiles shown by Consolandi et al. (2016) and to test the 
bar-quenching scenario proposed in Gavazzi et al. (2015). 
The radial color profile of barred galaxy shows that bars are on average redder than their surrounding disk producing an outside-in gradient toward red in correspondence
of their corotation radius. The distribution of the extension of the deprojected length of the bar suggests that bars have strong impacts on the gradients of averaged color profiles. 
The dependence of the profiles on the mass is consistent with the bar-quenching scenario, i.e. more massive barred galaxies
have redder colors (hence older stellar population and suppressed star formation) inside their corotation radius with respect to their lower mass counterparts.
 
 }

\maketitle


\section{Introduction}
Hydrodynamical simulations have made clear that bars have a major impact on the secular evolution 
of galaxies \citep{atha02,sell14}. 
It is well known that barred potentials exert non axisymmetric forces onto the gaseous component of the galaxy: the gas within the 
corotational radius is  
rapidly funneled to the center of the galaxy \citep[within the Inner Lindblad Resonance, see][]{KK04,K13,sell14, fanali15} while the gas outside is confined  
to the outer disk \citep{sanders76,shlo89,atha92,beren98, regan04,kim12,cole14}.
While it is not clear whether this phenomenon can trigger AGN activity or not \citep{emsell15}, it is out of question that the high density reached 
by the  gas dragged in the center of the galaxy triggers a burst of star formation that rapidly depletes it, turning into gas-poor
the region inside the corotational radius \citep{krum05,krum09,daddi10,genzel10}.
From an observational point of view, starting from the seminal work of E. Hubble who dedicated half of its world-wide famous 
tuning fork to barred disk galaxies \citep{hub36}, bars have increasingly captured 
the interest for the understanding of galaxy secular evolution.
As a matter of fact, throughout the years observations have enlighten the physical effects of bars 
on galaxies \citep{saka99,KK04,jo05,sh05,K13}
and, above all, the extremely high frequency of barred galaxies among spirals.
Indeed, among local bright disk galaxies, $\sim 60\%$ \citep{kna99,esk00,delmestre07,marinova07} are barred if observed in the near-infrared and about $\sim 40\%$ 
in the optical bands \citep{esk00,marinova07}, hinting at bars as fundamental drivers of the evolution of late type galaxies (LTGs). 

If and how the bar fraction evolves across the cosmic time is still under debate \citep{jo04,sh08}  
as well as the exact determination of the dependence of the bar frequency on stellar mass (especially at the faint end of the mass function) and
on galaxy environment \citep{thomp81,marinova12,skibba12,lans14,alonso14}.
For example,  \citet{masters12}, \cite{skibba12}, \citet{abreu12}, \citet{pg15} all consistently report a bar fraction that
increases with increasing mass.
Nevertheless, \citet{bara08} recover a strong-bar fraction that increases with decreasing mass while
\citet{nairbfrac} find a strong-bar fraction that decreases from $\sim 10^{9}$M$_\odot$ to $\sim 10^{10}$M$_\odot$ and increases again from  
$\sim 10^{10}$M$_\odot$ to $\sim 10^{11}$M$_\odot$. 

The work by \citet{cheung13} and \citet{pg15} underline the crucial importance of determining the exact dependence of the fraction
of strongly barred galaxies on total stellar mass. Namely, an increasing bar fraction with increasing mass would explain the 
central quenching of the star formation (SF) in high mass main sequence galaxies that bends the local 
star formation rate (SFR) vs stellar mass relation at  $\sim 10^{9.5}$M$_\odot$. Moreover, 
\citet{sanja10} report that below $\sim 10^{9.5}$M$_\odot$ disks are systematically thicker, making it difficult to develop bars.

\begin{figure}
\begin{centering}
\includegraphics[scale=0.65]{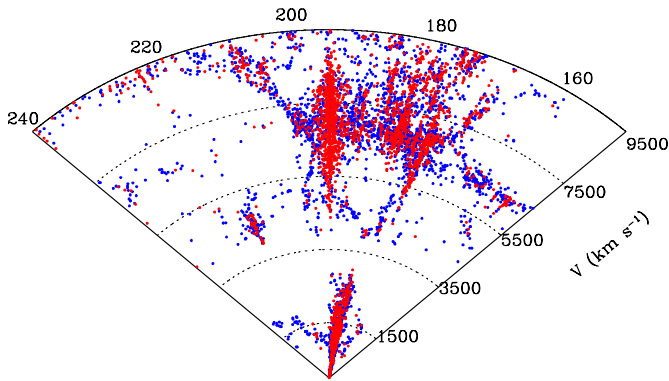}
\caption{Wedge diagram of galaxies belonging to the sample studied in this work: the Coma supercluster (c$z>4000$kms$^{-1}$) and the Local supercluster (c$z<3000$kms$^{-1}$). Blue
dots represent late-type galaxies wile red dots stand for early-type galaxies. }
\label{wedge}
\end{centering}
\end{figure}

\begin{figure}
\begin{centering}
\includegraphics[scale=0.4]{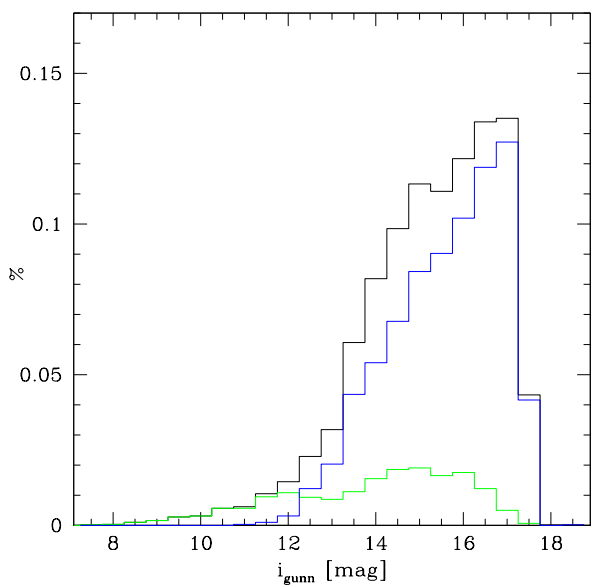}
\caption{Distribution of the $i$-band magnitudes of the whole sample and separately for the Local (green line) and the Coma (blue line) supercluster sample.}
\label{magdistr}
\end{centering}
\end{figure}

Determining robustly the real bar fraction below $10^9$M$_\odot$ demands statistics as well as sensitivity and resolution.
High redshift determinations lack of both aspects and
the best environment to determine the optical bar fraction in such a wide range of mass is therefore the local Universe, taking advantage of 
the publicly available data 
of the Sloan Digital Sky Survey \citep[SDSS,][]{sdss}. The SDSS fully covered the area of two nearby structures such as the Virgo and 
the Coma superclusters that give the opportunity to study thousands of galaxies down to a limiting mass as low as $10^7$M$_\odot$ with 
a physical resolution of $\sim 600$ pc at the distance of Coma.
However, the task is non-trivial, as stellar bars among the population of dwarf late type galaxies are often low surface brightness 
features that are easily confused with or hidden by poorly resolved patches of SF.
Purely visual inspection drives the classification of barred galaxies of most  of the catalogs of galaxies such as
 the cases of the RC3 \citep{rc3}, VCC \citep{vcc} and \citet{naircat} classifications.
Nevertheless, because of the subjective nature of visual inspection classification, it is habit to average the 
votes of as many as possible classifiers.
A clear example is given by the recent works by \citet{masters11}, \citet{masters12}, \citet{skibba12} and \citet{mel14} which are 
based on the classification of the Galaxy Zoo project \citep[{\it www.galaxyzoo.org}, see][]{zoo1,masters11}. 
This classification is made on the basis of thousands of  votes assigned by citizens that are asked to 
visually inspect SDSS galaxies and to answer questions about the morphology of the object.\\ 
On the other hand, a different approach is adopted for example by \citet{woz95}, \citet{jo04}, \citet{bara08}, \citet{bars09} and \citet{abreu12} who 
support the visual inspection by performing isophotal fitting analysis and looking at some precise features in the ellipticity and 
position angle (P.A.) radial profiles.

In this paper is tested a different method to produce an objective classification that does not rely on visual inspection of neither the images 
nor the ellipticity/P.A. radial profiles of galaxies and extract a fairly pure sample of barred LTGs.  
An automatic procedure based on the sample and on the methods explained in \citet{C16}
for the isophotal fitting (explained in section \ref{tilt}) ran on $\sim6000$  SDSS galaxies in the Local and 
Coma superclusters. 
Throughout section \ref{vote} is presented the automatic 
bar-extraction criteria based on the ellipticity and P.A. profiles extracted.
Limitations, pureness and completeness of the automatic classification are discussed in section \ref{check} by comparing the final 
selection to other classifications found in the literature. 
In section \ref{bfrac} the resulting bar fraction vs mass relation is compared to other published results.
Finally, the results are discussed in section \ref{sunto} and 
the bar-quenching scenario described by \citet{pg15} is tested by creating a set of templates of color profiles of barred galaxies 
in different bins of mass. These will be also compared to the template color profiles produced in \citet{C16} (from now on C16).

\section{The sample}
The sample analyzed in this work  is displayed in Fig. \ref{wedge} and coincides with the one presented by C16.
It comprises 6136 nearby galaxies in the spring sky selected from the SDSS and further split in two subsamples: 
i) the Local Supercluster ($11^h<RA<16^h$; $0^o<Dec<18^o$; $cz<3000 ~\rm km ~sec^{-1}$) 
containing 1112 galaxies and includes the Virgo cluster; 
ii) the Coma Supercluster ($10^h<RA<16^h$; $18^o<Dec<32^o$; $4000<cz<9500 ~\rm km ~sec^{-1}$) containing 5024 galaxies
and includes the Coma cluster. 
Since galaxies at the distance of Virgo have apparent size often exceeding 5 arcmin, they are strongly affected by 
the shredding problem (Blanton et al. 2005) therefore our catalog cannot solely rely on the SDSS spectroscopic database. 
Briefly, the Local supercluster sample is selected following the prescriptions of \citet{pg12}:
in the area occupied by the Virgo cluster, the selection is based on the VCC catalog (limited however to $cz<3000 \rm
~km ~sec^{-1}$) down to its magnitude completeness limit of 18 mag \citep{vcc}. The object selection is furthermore limited 
to objects with surface brightness above the $1\sigma$ of the mean sky surface brightness in $i$ band 
of the SDSS data (C16). Outside the Virgo cluster the SDSS selection is complemented with objects taken from NED and ALFALFA \citep{alfa40}. 
At the distance of the Coma supercluster instead the shredding problem is less severe and therefore we followed the
selection of \citet{pg10, pg13b}: galaxies are selected from the SDSS spectroscopic database DR7 \citep{dr7} with $r<17.77$ mag and 
to fill the residual incompleteness of the SDSS catalog for extended galaxies and due to fiber conflict,  
133 galaxies from the  Catalog of Galaxies and Clusters of Galaxies  \citep[CGCG, ][]{cgcg} with known redshifts from NED and 28 from ALFALFA are added, reaching a total of 5024 galaxies.

For these galaxies, C16 downloaded the SDSS images \citep[using the on-line Mosaic service,][]{montage} only
in the $g$ and $i$ band for mainly two reason:
$i$) their higher signal-to-noise ratio (SN) compared to the u and z filters, 
$ii$) the $r$ filter have a central wavelength closer to the $g$ filter central wavelength with respect to the $i$ filter, making 
the $g$-$i$ color more sensitive to stellar population gradients.
The download process worked in both the $i$ and the $g$ band for 5753 (94\%) targets out of which, 221 galaxies were discarded a posteriori because they lie too close 
to bright stars or they have too low surface brightness.

The remaining analyzed sample is constituted of 5532 galaxies that can be considered representative of the nearby universe. 
The distribution of the $i$-band magnitudes of this sample is displayed in Fig. \ref{magdistr}.
Galaxy masses are computed assuming a Chabrier IMF, following \citet{zibi09} from the $i$-band luminosities of the galaxies and their ($g$-$i$) color
published in C16. 

\section{Automatic tilted profiles for non axis-symmetric structures detection}
\label{tilt}
C16 presented a semi-automated IDL procedure that performs photometry and extract surface brightness and color profiles
over multi-band SDSS images of local galaxies.
Briefly, stamp images ($g$ and $i$ band) centered on target galaxies are downloaded from the on-line Mosaic service \citep{montage}.
The procedure automatically performs sky-subtraction in each filter and creates an averaged white image with an improved signal-to-noise ratio.
This image is processed by Source Extractor \citep{sex96} which detects stars and galaxies producing a map of the fore- and background
objects that is exploited to create an accurate mask of the field of view. 
Simultaneously, the external geometry of the galaxy is extracted by Source Extractor and exploited by our routine to  
create (in the white frame) a set of concentric ellipses fitting the target and enabling the evaluation of the surface brightness profile 
in the $g$ and $i$ images as well as the $g$-$i$ color profile. 
The latter are therefore extracted using ellipses with fixed ellipticity and position angle and
do not retain any imprint of a possible bar component, such as the twist of the isophotes or a peak in the ellipticity profile.\\
In this work, I describe the implementation of an accessory IDL-based routine of the C16 procedure able to quickly perform ellipse fitting \citep{jo04,marinova07,bars09} and
automatically extract the barred galaxies among the sample. 
This tool makes use of two fundamental outputs of the C16 procedure: the mask of fore- and background objects, the white light image and the white light surface brightness profile.
Eventually, the new routine  returns for each galaxy the ellipticity and P.A. of the fitted ellipses as a function of the semi-major axis 
and the surface brightness and color profiles derived from the same ellipses.
\begin{figure*}
\begin{centering}
\includegraphics[scale=0.68]{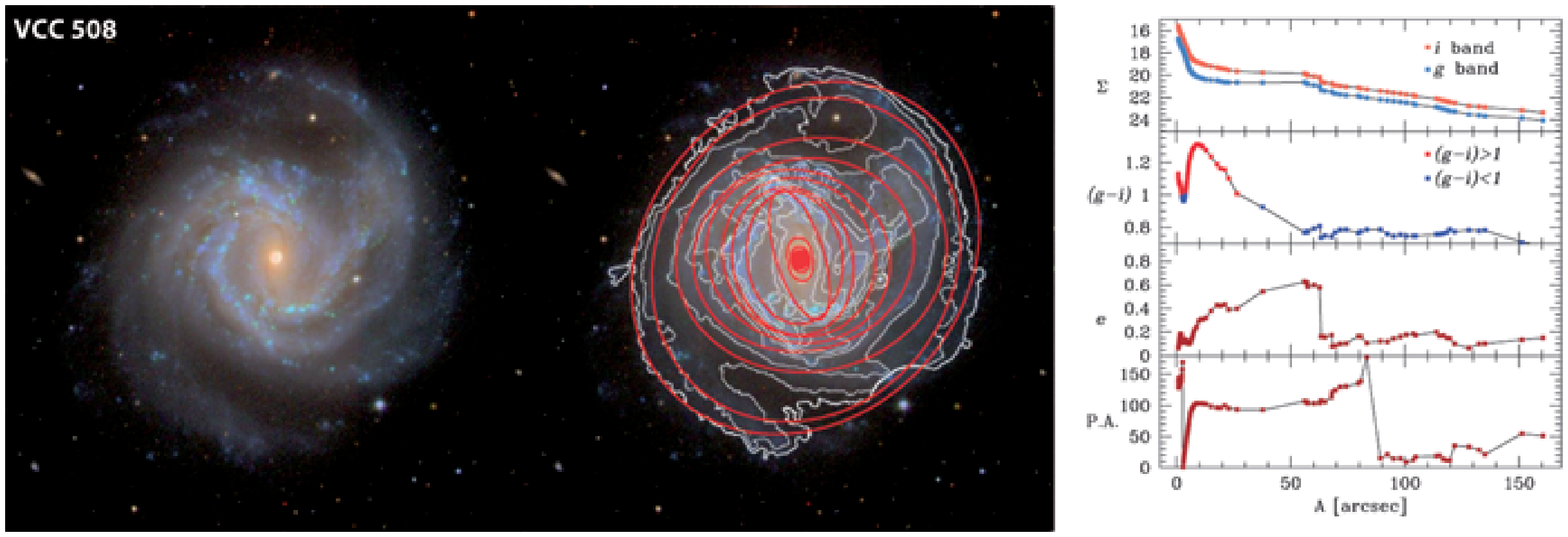}
\includegraphics[scale=0.68]{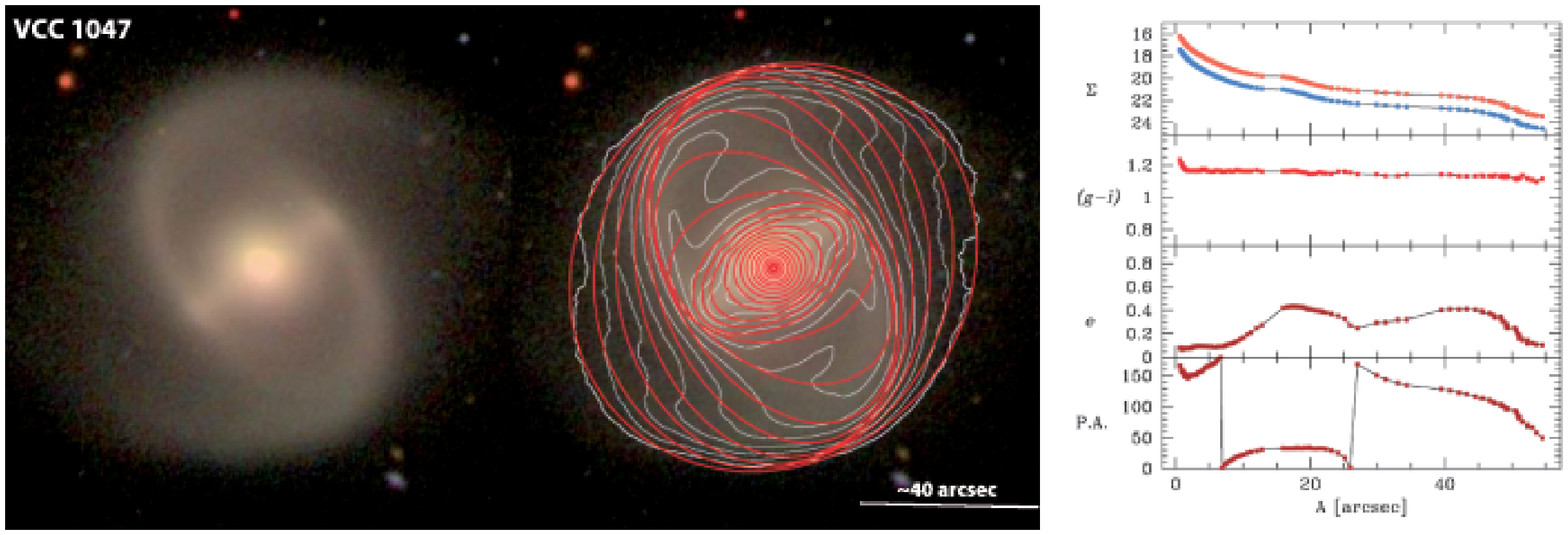}
\includegraphics[scale=0.68]{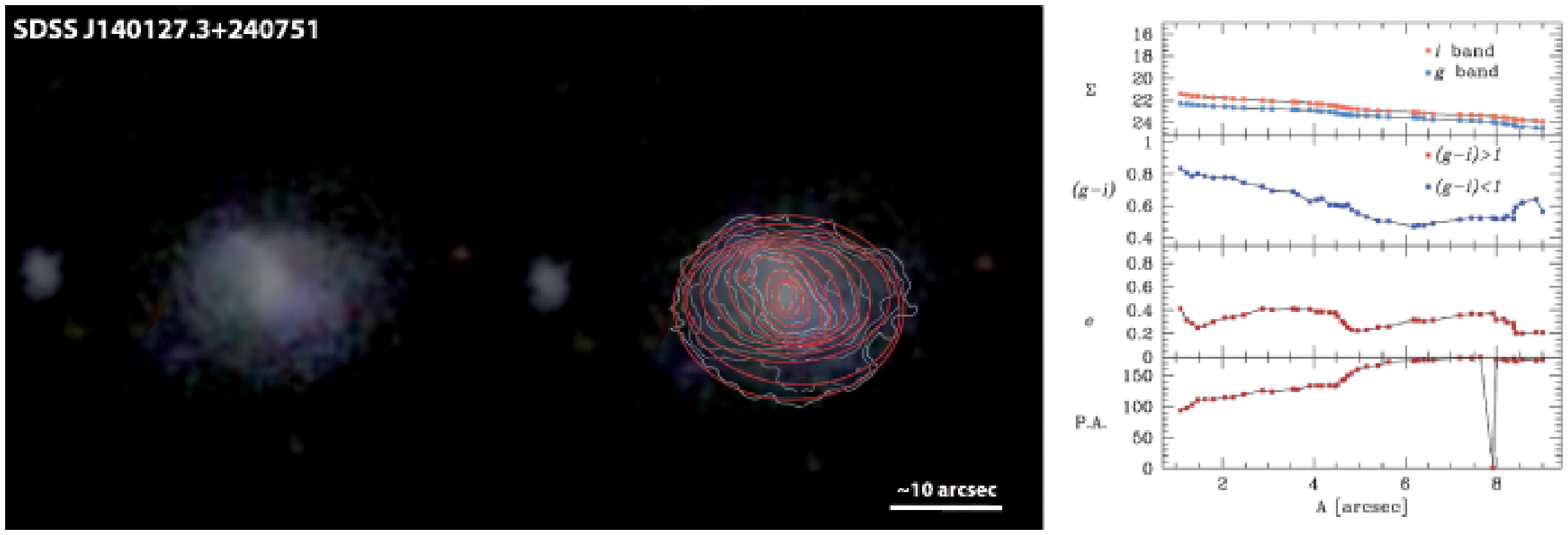}
\includegraphics[scale=0.68]{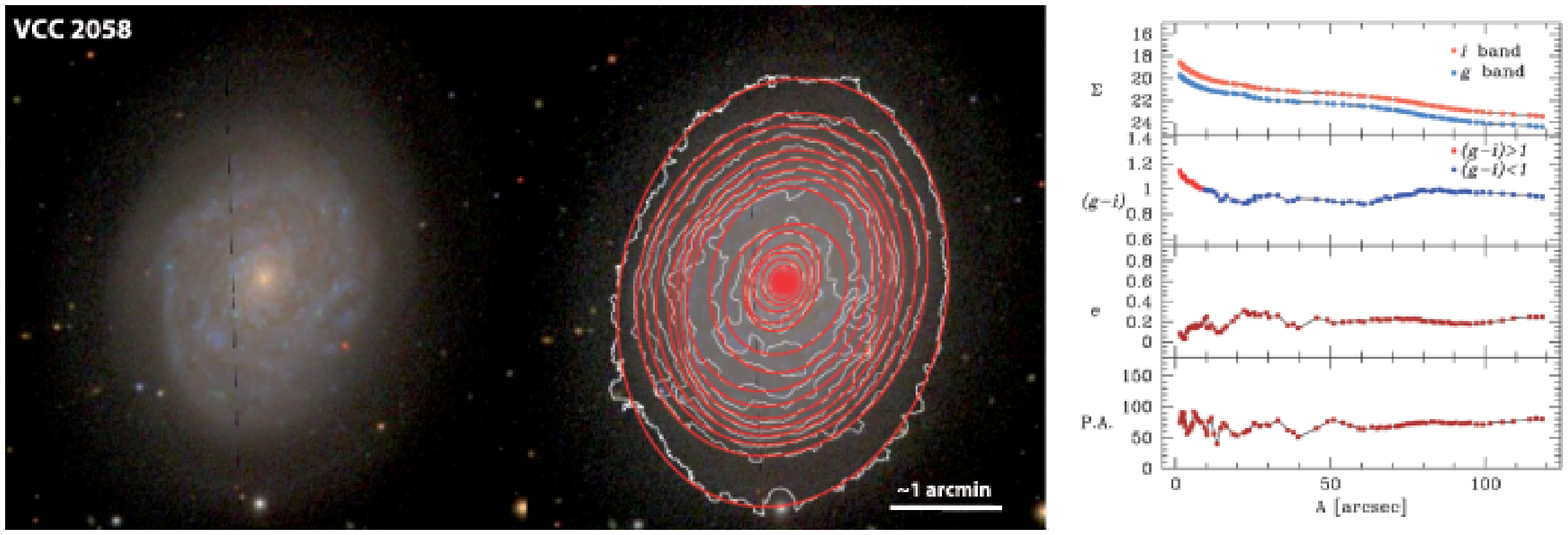}
\caption{Images of three barred galaxies and one unbarred, respectively VCC-508, VCC-1047, SDSS-J$140127.3+240751$ and VCC-2058. 
For each object, I plot the surface brightness and the color profile
as well as the ellipticity and P.A. of the fitted ellipses as a function of the semi-major axis.  }
\label{profil}
\end{centering}
\end{figure*}

\subsection{ellipse fitting}
In order to correctly evaluate color profiles, surface brightness profiles must be extracted over consistent apertures in both the $g$ and $i$ band.
The ellipse fitting procedure is performed on the white image which represents a high signal to noise reference frame common to all filters.
For each galaxy, the procedure fits ellipses to the contours of the white light frame, that are built owing to the {\it cgContour} routine of the IDL 
Coyote Library.
The routine is set to extract 100 logarithmic levels ranging from $3\times \sigma_{sky}$ to the maximum of the white surface brightness (non-tilted) profile, 
previously evaluated by the procedure described in C16.  
The {\it cgContour} routine returns the two-dimensional distribution of pixels of each contour level in the white frame.
Starting from outside the galaxy and going toward the center, 
the procedure automatically fits each contour with the parametric formula of an ellipse rotated of a position angle (P.A.)\footnote{Each ellipse is fitted owing to the {\it mpfitellipse}
procedure that is based on the {\it mpfit} routine, a non-linear least squares fitting program described in \citet{mpfit}.}.
For each contour level I setup the initial guesses for the ellipse-fitting assuming the center X$_c$, Y$_c$ of the ellipse as the average $X,Y$ pixel coordinates composing
the contour level distribution. Further on the initial guesses for the semi-major (-minor) axis are chosen as half the distance between the first (/half-path) pixel coordinates 
and the one at half-path (/three quarters) while the initial guess for the P.A. is taken as 0.
Each fitted ellipse must not overlap the previously fitted ellipse.
Once all contours are fitted, the routine checks and discards ellipses that are not centered.
To this aim, the coordinates of the center previously evaluated by Source Extractor are not considered and instead recomputed
as the mode of the coordinates of the center of all fitted ellipses \footnote{ This choice was made because the coordinates 
evaluated by Source Extractor are referred to the ellipse that fits the external (at the $1.5\Sigma_{sky}$ isophote) geometry 
of the galaxy which does not always provide the correct barycenter of the internal isophotes}.\\
The procedure discards ellipses whose center is at a distance from the galaxy center greater than $\sim 15$ arcsec for Virgo galaxies and $\sim 5$ arcsec in Coma.
Moreover, in some cases, especially among irregular galaxies, the most internal ellipses fits contours that follow
patches of star formation not centered on the galaxy but still within the tolerance distance adopted.
To fix this, ellipses that have central coordinates at a distance to the galaxy center that is greater than their semi major axis ({\it sma}) are 
not considered.\\
Finally, the surface brightness of each ellipse in the $g$ and $i$ band images are computed
measuring the average surface brightness along the path of each fitted contour and $(g-i)$ color profiles are consequently obtained making the 
difference between the $g$ and $i$ surface brightness profiles. 
Examples of the ellipses fitted and of the profiles extracted by the procedure are shown in Fig. \ref{profil} for four different galaxies 
(three barred and one unbarred).
\subsection{bar selection}
\label{vote}
As it is described by \citet{woz95}, \citet{jo04}, \citet{marinova07}, \citet{bara08}, \citet{bars09} and as it can be seen in  Fig. \ref{profil}, bars in 
face-on galaxies produce a peak in the ellipticity profile to which a plateau in the P.A. is associated, i.e. the P.A. is constant within $\pm 20^o$.
The ellipse fitting method has been extensively tested to be efficient in extracting barred galaxies among wide samples by many authors
\citep{jo04,marinova07,bars09,lauri10}.
It has two free parameters, $\Delta e$ and $\Delta P.A.$. The first one is the difference between the ellipticity of the bar and the one of the disk while the second
parameter is the interval within which the P.A. varies along the bar.\\
The procedure analyzes the properties of the ellipticity and P.A. profiles of each galaxy and hunts bars automatically extracting those objects that 
exhibit $\Delta e>0.08$ and $\Delta P.A.>20^o$. These threshold values were tested by \citet{bars09} who showed that these values minimize
the fraction of spurious/bad detections and simultaneously maximize the bar identifications.
In order to automatically select galaxies on these basis, the procedure has to correctly identify the ellipticity peak 
and the related P.A. plateau associated to the bar.
Obviously both variations of the ellipticity and P.A. are more evident and easier to detect in face-on galaxies, where the contrast between the geometry of 
the galaxy and of a bar is maximum. 
Hence we limited the extraction to galaxies with $B/A$>0.7, where the axis ratio $B/A$ is evaluated by the 
procedure itself as the average ratio  between the semi-minor (B) and semi-major (A) axis  of the four most external ellipses fitted. 
The blind extraction happens in three distinct phases: 
in the first phase (i), the procedure finds each ellipticity peak in the ellipticity profile and looks 
for each of them for a related plateau in the P.A. profile; 
in a second instance (ii), the ellipticity peaks associated with a plateau are compared between them and the most promising peak is extracted; 
finally (iii), the geometry of the isophote associated to the extracted peak is compared to the galaxy geometry and the final word on the possible bar presence
is spoken.\\
Step (i): In order to identify the plateau, for each peak the routine considers all fitted isophotes in the neighborhood of 
each considered peak with  P.A. within $20^o$ from P.A.$_{peak}$ an ellipticity within 0.1 from $e_{peak}$. \\
Step (ii): Peaks found in (i) are all thought to represent a possible bar and are therefore put into competition.
Each peak receive a positive or negative vote according to their ellipticity, $\Delta e$, length of the plateau, number of fitted 
ellipses. In other words, the procedure gives the best votes to the peaks with the greatest ellipticity that: 1)maximizes the ratio between the length 
of the plateau and the {\it sma} at which the peak is found; 2) minimizes the variations in the P.A. profile; 3) is found in 
the region $0.05<a_{peak}/a_{gal}<0.95$.
Eventually only the best ranked peak is considered and if two (or more) peaks get the same vote, the preference is given to the most internal peak
\footnote{As a matter of fact, spiral arms can produce a peak in the ellipticity profile and a P.A. plateau that mime the behavior of a bar but, 
when the bar is present, they have larger scale length.}.\\ 
Step (iii): once that a single-peak is extracted, its vote is further boosted according to its values of $\Delta e$ and $\Delta P.A.$, i.e. 
a peak that has $\Delta e \sim 0.4$ and $\Delta P.A. \lsim 10 $ will receive a higher rank compared to a peak that has $0.1 \lsim \Delta e \lsim 0.2$ 
and $10 \lsim \Delta P.A. \lsim 20$.\\

\section{Selected bars}
\label{check}
In order to account only for systems that can develop non-axisymmetric structures in the barfraction estimate, a morphological cut is applied to exclude
pressure supported systems. The morphological classification of all the galaxies of the sample is found in the online public database GOLDMine \citep{pg03,pg14b}. 
Nevertheless this classification is purely visual \citep{vcc} and it
is likely not accurate at separating slowly rotating systems (elliptical galaxies) and fast rotators (namely S0s), 
preventing from a robust estimate of the bar fraction when including early-type disks. 
Therefore I will give the results separately for the 
sample of late-type disks (from S0a to Sm and Irr) and the sample that includes S0s.
The procedure performed the extraction over a subsample of 922 face-on ($B/A>0.7$) late-type galaxies (LTGs, from S0a to Sm and Irr) and over 447 S0s and dS0s. The 
sample including early type disks therefore accounts for 1365 members.
In the late-type disks sample
365 galaxies received a positive vote (442 including early types), thus have been highlighted to be possible bars.
A visual inspection of all galaxies revealed that above vote$=4$ (227 galaxies) the sample of bars extracted can be considered more than $95\%$ pure while below 
$vote=1$ (557 galaxies) only $\sim 20$ barred galaxies were erroneously classified as unbarred ($3-4\%$).
The $votes=$2,3,4 categories (106) comprehend about $10\%$ of bad detections and $\sim 9\%$ of ambiguous cases in which, 
even after visual inspection of both images and profiles, it is very hard to speak a final word on their real morphology. This is summarized in Tab. \ref{tab1}
that indicates that selecting galaxies with $vote\geq2$ defines a satisfactory $\sim 90\%$ (considering as intruders both bad and uncertain cases) pure sample of barred galaxies that maximizes 
the overall bar fraction of our sample ($36\%\pm2\%$).

\begin{table}[t]
\caption{Frequencies of bad, uncertain and  confirmed detections after visual inspection of the 922 face-on late-type galaxies.
Bad detections are: the missed bars, in the category having vote lower than 0, or, in the other groups, the unbarred galaxies considered as barred.}
\centering
\begin{tabular}{c c c c}
\hline\hline
                 & N$_{gal}$   & Bad classification & 	Uncertain  \\
\hline\hline
 vote $\leq$ 0	 &	557    &	22	    &    10 	   \\
 vote $>$ 0	 &	365    &	30	    &    25	   \\
 vote = 1	 &	32     &	10	    &    4	   \\
 vote = 2	 &	24     &	3	    &    4	   \\
 vote = 3	 &	66     &	5	    &    4	   \\
 vote = 4	 &	16     &	2	    &    2	   \\
 vote $\ge$ 2	 &	333    &	18	    &    21	   \\
 vote $\ge$ 4    &	227    &	10	    &    11	   \\
 
  \hline
\end{tabular}
\label{tab1}
\end{table}

\subsection{Comparison with visual classifications}
To further test the efficiency of the extraction of barred objects I compared this classification to the one performed by Binggeli et al. (1985)
for galaxies of the Virgo Cluster Catalog (VCC). 
Among the VCC, face-on LTGs having a $B/A>0.7$ and a $cz<3000 kms^{-1}$ are 56, out of which 20 are classified as barred.
Among VCC bars, $80\%$ have been consistently classified as barred also by my automatic method which, on the other hand, classifies as barred 16 more galaxies
(among which there is VCC 508, see Fig. \ref{profil}a) and misses 4 barred galaxies from the VCC implying an overall accordance between the present classification and Binggeli of $\lsim70\%$.
Nevertheless it is worth stressing that, among the objects classified as barred  by the automatic pipeline and unbarred by the VCC, 50\% are instead classified as barred in the RC3
\citep{rc3}, indicating that the bar extraction efficiency of the pipeline is better than what we can deduce from the accordance with VCC.
The remaining $50\%$ (9 galaxies) are either classified as peculiar objects or weakly barred: this are mostly low surface brightness systems with some degree of perturbation and asymmetry.
Only 4 galaxies do not show any bar component and are classified as normal late type galaxies by both RC3 and VCC. \\

A more recent classification based on SDSS images has been performed  over a sample of about 500 galaxies
in the Virgo cluster by \citet{abreu12} and over a different sample of about 200 galaxies in the Coma cluster by \citet{abreu10}.
I found a sample of 452 objects in common with the sample of this work and the ones of \citet{abreu12}.
Performing our selection cut in their sample leads to a test sample of 29 galaxies out of which 14 are barred in the \citet{abreu12} classification.
Among these, 12 have been consistently voted as barred galaxies by my automatic procedure while 2 galaxies where erroneously considered unbarred and other 2
that are not classified as barred in the \citet{abreu10,abreu12} classifications. 
If we instead include the analysis of S0s we find a test sample of 50 galaxies. In this case \citet{abreu10} finds 23 barred objects while the automatic procedure extracts a total of 
18 bars. Overall, 17 galaxies were classified as barred by both works while 6 galaxies where missed by my automatic procedure. The overall agreement in the classification of 
the 50 objects selected is $\sim 82\%$.

Finally we compare the classification with the one extracted in the Galaxy-zoo \citep{zoo1,zoo2} selecting objects with pbar > 0.5 and that have been vote by more
than 28 citizens \citep{mel14}.
In total I found a Galaxy-zoo (GZ) classification for 727 late-type galaxies with $BA>0.7$ out of which 180 barred ($25\%$).
On the contrary our procedure finds 285 barred galaxies ($\sim 39\%$) including $80\%$ of the galaxies classified as barred by the zoo.
The objects that are not found barred in the GZ project are mostly ($62\%$) galaxies with M$_* < 10^{9.5}$ M$_{\odot}$. 
In these cases, the procedure is sensitive to the non axisymmetric distribution of bright HII regions and clumps in these irregular 
galaxies leading to wrong or ambiguous detections that overestimates the fraction in the low mass systems.
On the other hand, among brighter galaxies, it was possible to find an independent classification of the RC3 and/or VCC catalogs indicating as barred
$\sim 30\%$ of these objects  voted unbarred in the GZ, indicating the good work of the procedure at higher luminosities. Moreover about $50\%$ of
the remaining galaxies have a pbar between 0.35 and 0.5 indicating that they were nearly recognized as barred in the GZ.
Therefore there are evidences that the present procedure overestimates the fraction at low mass but performs quite good at higher masses for an overall
agreement with this classification of $\sim 78\%$.

In spite of the aforementioned difficulties of this automatic method at low mass, the overall agreement among the classifications is 
satisfactory but with an high source of uncertainty coming from bars detected in low mass systems. 
 These are often low surface brightness with poorly
resolved structures that, even after visual inspection of their images and ellipticity profiles, it is very difficult to say if they harbor bars (see Fig.
\ref{profil}, the galaxy SDSSJ140127.3+240751).
\begin{figure}
\begin{centering}
\includegraphics[scale=0.45]{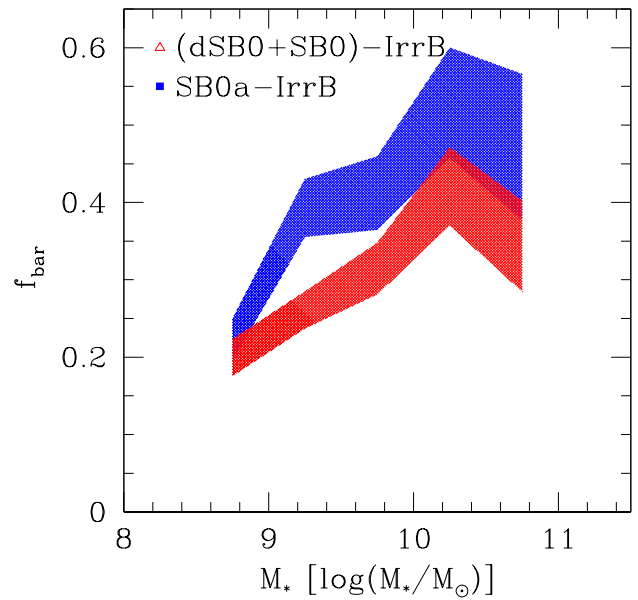}
\caption{Optical local bar fraction as a function of stellar mass given separately for the sample that includes galaxies from S0a to Irr (blue) and
the sample that accounts for SB0s (red). The width of the shaded area gives the poisson uncertainty in each bin.}
\label{bfr}
\end{centering}
\end{figure}

A further demonstration comes from the comparison with the visual classification performed in Gavazzi et al. (2015) by all the authors.
In total, I could compare the classification of 229 galaxies out of which 51 were classified as barred by the authors.
In this work the procedure was able to spot  $\sim 80\%$ of the bars classified in Gavazzi et al. (2015) but, on the other hand, the 
procedure classifies 60 more bars.
Of these, the $68\%$ are once again low-mass (M$_*$<$10^{9.5}$M$_{\odot}$) irregularly shaped systems.
Of the remaining more massive galaxies that were not classified as bars by \citet{pg15}, $33\%$ is confirmed to host a bar by other 
independent classifications found in NED that corroborate the result of this method. Once again, the procedure shows a good capability in spotting bars in high 
luminosity systems while having more difficulties at low mass. 

\section{The local bar fraction }
\label{bfrac}	  
In the past section we described our selection
and showed that the procedure has selected a reliable sample of barred galaxies.
C16 showed that spiral galaxies above some threshold mass are undoubtly redder 					   	     
then their lower mass counterparts and that this phenomenon is more evident in their central parts.
I will now test the consistency of the evaluated bar fraction with the ones extracted by \citet{abreu12} and Gavazzi et al. 2015
and check if the presence of a bar can indeed produce the trend in color profiles with mass that C16 observed by 
building a template color profile of barred galaxies.

The overall local bar fraction evaluated in the present analysis is $36\%$ while if we include S0s it becomes $28\%$.
The agreement with \citet{jo04}, \citet{marinova07}, \citet{bara08}, \citet{nairbfrac}, \citet{pg15} is satisfactory.
In Fig.\ref{bfr} is shown the local bar fraction  as a function of mass obtained separately for late-type galaxies and late-type galaxies plus S0s and dS0s.
This has been evaluated in 5 bins of 0.5 dex from $10^{8.5}$M$_{\odot}$ to $10^{11}$M$_{\odot}$. 
The bar fraction increases evidently with increasing mass in both relations 
confirming many literature results \citep{abreu12,masters12, skibba12,cheung13, pg15}. 
Nevertheless there is a clear separation between the sample containing lenticular galaxies and the LTG-only sample.
Compared to the bar fraction published by \citet{pg15}, the fraction at low mass is considerably higher. 
As discussed in the previous section, the automatic pipeline is very sensitive to irregular structures in low mass galaxies and may produce 
an higher fraction at low mass. In this sense, the low mass end of this bar fraction could be considered as an upper limit. Nevertheless, the 
steep relation between the total stellar mass and the bar occupation fraction is preserved.
\begin{figure}
\begin{centering}
\includegraphics[scale=0.45]{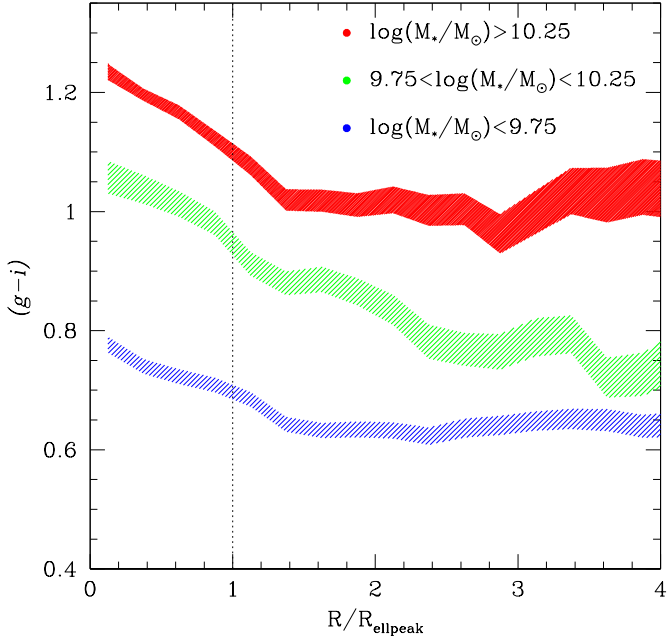}
\caption{Color profiles templates of barred galaxies automatically extracted in this work in three bins of increasing mass.}
\label{tmpl}
\end{centering}
\end{figure}
\begin{figure}
\begin{centering}
\includegraphics[scale=0.45]{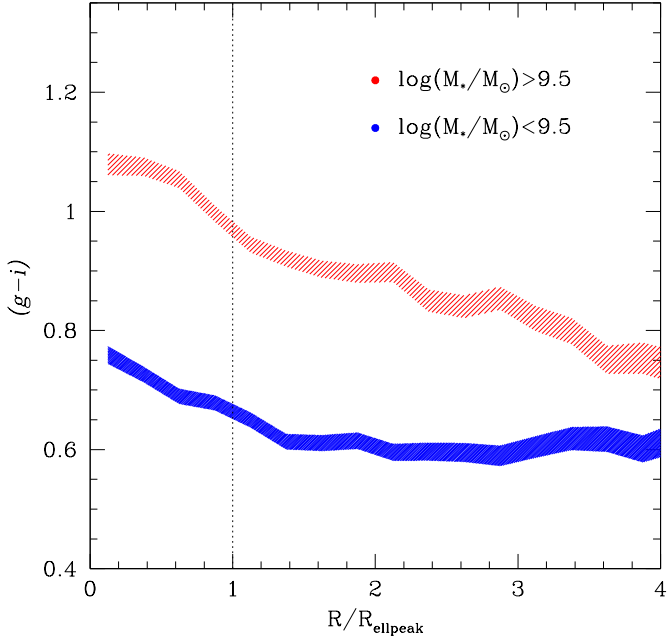}
\caption{Color profiles templates of barred galaxies automatically extracted in this work in two bins of increasing mass: above (red) and below (blue) M$_{\rm knee}$, the 
threshold mass indicated by Gavazzi et al. (2015).}
\label{mkn}
\end{centering}
\end{figure}
\section{The colors of bars: discussion and conclusions}
\label{sunto}

Holding the selection onto late-type galaxies and excluding S0s, I constructed a template $(g-i)$ color profile in three different bins of mass:
i)  below $M<10^{9.75}$M$_{\odot}$ ii) $M>10^{9.75}$M$_{\odot}$ and $M<10^{10.25}$M$_{\odot}$ iii) $M>10^{10.25}$M$_{\odot}$.
The bins were selected  to guarantee that more than 50 galaxies contribute to each template profile.
Color profiles are good tracers of the specific SFR and works such as \citet{mac04,mcd11}, C16 have highlighted the good 
correspondence between average age of the stellar population as a function of radius and color radial profiles.
Moreover, using the technique of template profiles, C16 have shown that massive spiral galaxies develop a red and dead component, the importance
of which increases with mass. Using the same technique on the tilted color profiles of the 
subsample of barred galaxies of C16 extracted in this work, we 
test if the central red and dead component is consistent with the presence of a bar-like structure in the center of galaxies.
The template profiles for barred galaxies are displayed in Fig.\ref{tmpl} with radius normalized
to the radius of the selected ellipticity peak (R$_{ellpeak}$), a good proxy for the bar length and corotation radius \citep{lauri10}, with a radial step of
0.1 $\rm R/R_{ellpeak}$.
\begin{figure*}
\begin{centering}
\includegraphics[scale=0.45]{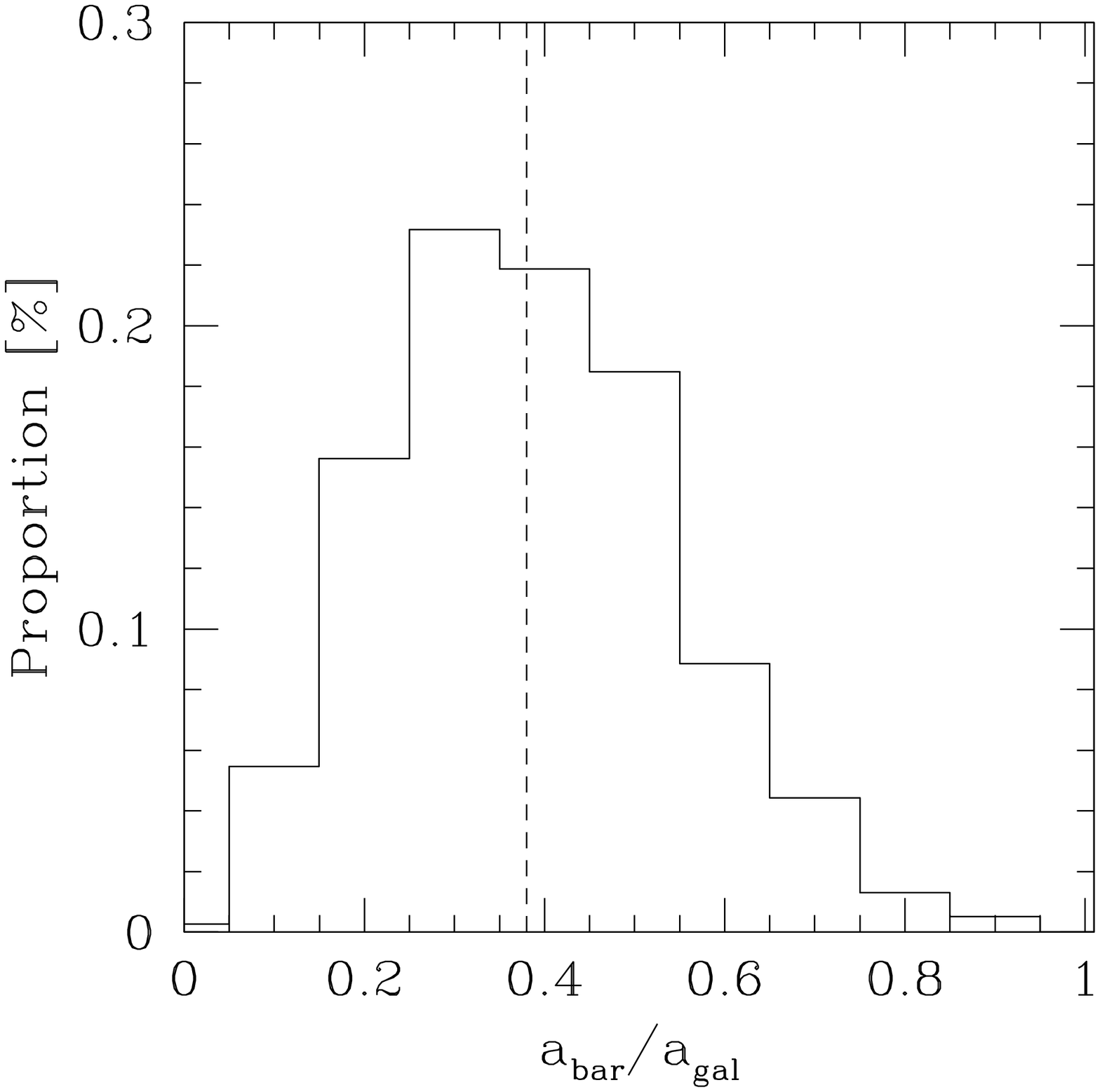}\includegraphics[scale=0.45]{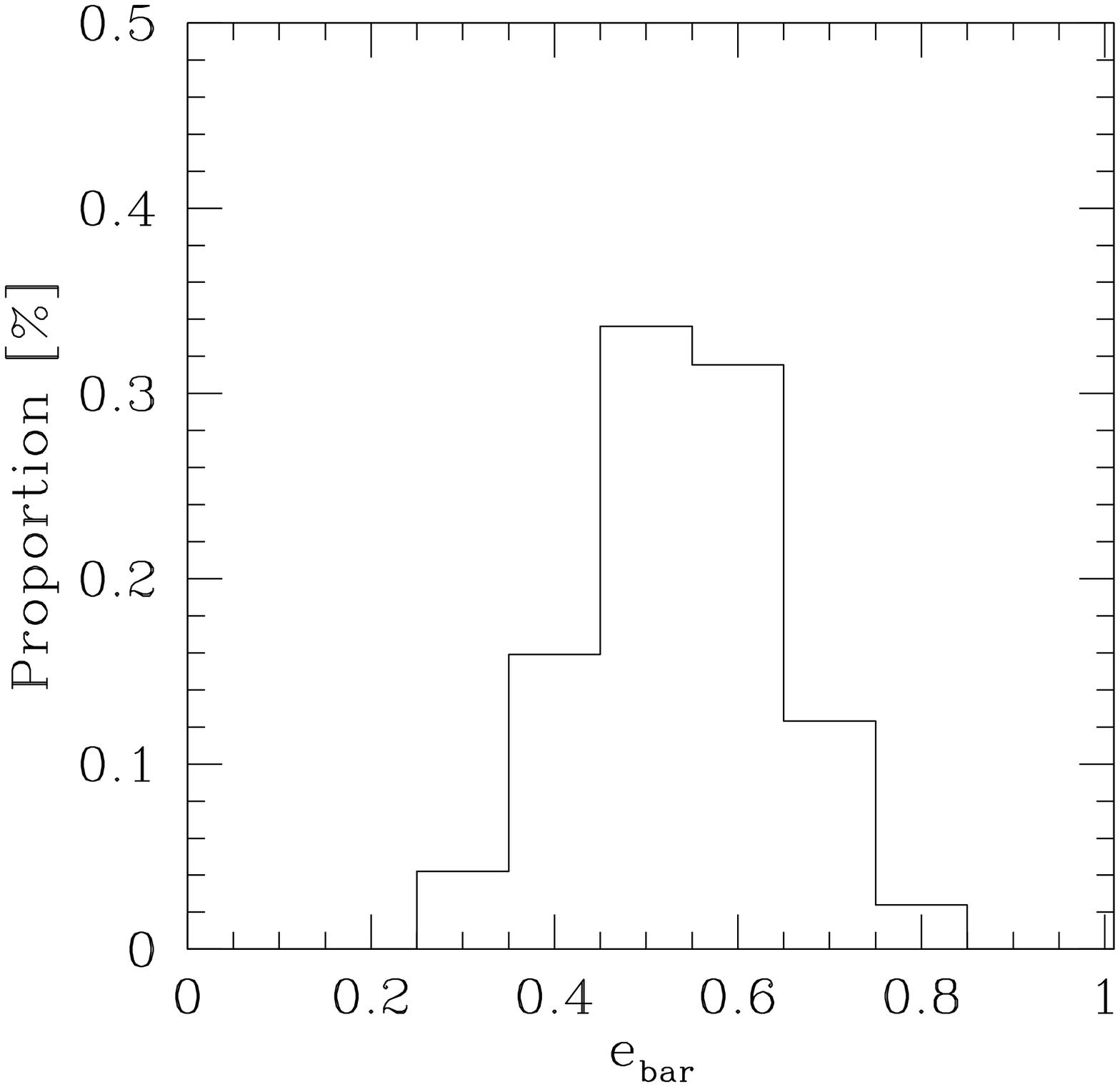}
\caption{(left) Distribution of the ratio between the deprojected semi major axis of the bar and the semi major axis of the galaxy. The dashed line indicates the average
value of the distribution equal to 0.38.
(right) The distribution of the ellipticity of the bars extracted in this work.}
\label{barextent}
\end{centering}
\end{figure*}

From low to high mass, the template profiles evolve significantly.
In the lowest mass bin the color profile is blue over all radii with possibly only a mild gradient toward red inside the bar radius.
Things change clearly in the intermediate mass template profile: the outer disk (R/Rpeak > 1) is blue and in the outermost regions overlaps with 
the lowest mass bin profile, while inside the corotation radius (R/Rpeak <1) the color profile is red as an elliptical of the same mass.
The bar has already reached the red sequence and, on the contrary, the disk is still on the blue cloud.
In the highest mass bin the red component has once again the typical color of the red sequence in the same range of mass. 
The disk is still bluer in the outer region but displays an average color of the disk that is redder than the respective 
average blue cloud values of the same mass.
Moreover in Fig.\ref{mkn} we show the template radial color profiles of barred galaxies for two samples up and below the threshold mass indicated by \citet{pg15}
to be the mass above which bars have the region inside the corotation radius quenched and therefore red. 
In order to quantify the average extension of the region that under the bar influence we correct the radius of the ellipticity peak 
for projection effect using the measured P.A with respect to its galaxy that is considered to have the P.A. of the last fitted isophote.
The distribution of the ratio of the deprojected bar semi major axis and the galaxy semi major axis is plotted in Fig \ref{barextent}. 
This distribution peaks at 0.3 consistently with others results such as the one published in \citet{marinova07} and \citet{bara08}.

The bars that we extracted are primarily strong bars \citep[e$> 0.4$,][]{lauri10} and  weak bars represent only the $\sim 9\%$ of the bars extracted
which is again consistent with the proportion observed in \citet{marinova07}.

Looking at Figure \ref{tmpl} we can deduce that bars are on average redder structures if compared to their associated disks.
Fig \ref{bfr} reveals that especially at high mass bars are extremely common and will likely have 
a big impact on the average photometric properties of the galaxy population.
Therefore bars will be strong contributors of the trends in color shown in the template color profiles of C16 
especially for high mass objects.
Moreover a further clue comes from the distribution of the ratio between a$_{bar}$ and a$_{gal}$ (Fig. \ref{barextent})
indicating that the average optical extension of the 
bar is $\sim 0.3$ a$_{gal}$ which is consistent with the extension of the intermediate/internal zone identified
in the average color radial profiles of C16 that is on average redder than the outer disk zone. 

A possible explanation for such a correspondence between 
the presence of the bar and the color of the galaxy, is the one proposed by \citet{pg15} who finds that 
the sSFR of main sequence local galaxies have a downturn at high mass indicating that massive disks have suppressed 
sSFR with respect of their lower mass counterparts. These authors suggest that the torque exerted onto the gaseous component 
by the bar, funnels the gas inside the corotation radius to the very center of the galaxy where it's rapidly consumed by a burst of star formation.
The region within the bar extent is therefore gas depleted and grows redder with time and this phenomenon can occur earlier in more massive 
disks which are dynamically colder. On the contrary, the gas outside the corotation radius is hold in place and keeps feeding the star formation
maintaining the disk blue.
Nevertheless the bar fraction versus mass relation, along with the well known color-mass relation \citep{C16}, implies that there is an higher fraction of bars 
among more massive galaxies with redder total colors, although we stress that these are still star forming spiral galaxies. This is still consistent with
previous works such as the one of \citet{masters11}, \citet{alonso13}, \citet{alonso14} who consistently find an increasing barfraction in redder galaxies.

I found a difference between the bar fraction evaluated among all LTGs (ty>1) and the one that embraces lenticulars.
As a matter of fact when lenticulars are taken into account the bar fraction decreases at all masses.
At this point, a note of caution is required: the morphology selection relies only on visual morphological classification \citep{vcc} 
which cannot disentangle the small population \citep[$\sim13\%$ of ETGs,][]{a3d,emsell11} 
of slow rotators (pure ellipticals) from the much wider population of fast rotators (disks that should have been taken into account when calculating
the bar fractions). Therefore this estimate of the bar fraction of the joint population of LTGs and S0/dS0s could be 
biased by the morphological classification. Nevertheless, the proportion that stands between fast and slow rotators \citep[respectively $\sim87\%$and $\sim13\%$ of ETGs,][]{emsell11} 
implies that the bar fraction could be even lower, as many fast rotators have been likely classified as ellipticals and therefore
excluded from the bar fraction determination.
The lower bar fraction can possibly arise from two different scenario:
i) S0s are older systems with respect of other disk galaxies and have already undergone buckling instability that weakened the bar;
ii) given the well known density-morphology relation \citep{dress80}, S0s populate dense environments which prevents from growing bars because of tidal interactions
or fast encounters.
Nevertheless it is worth stressing that other results suggest that intermediate/high density environments such as groups can
indeed enhance the possibility of growing a galactic bar \citep{skibba12}. 

As a final note, I would like to highlight that in the highest mass bin the bar fraction is lower. This feature has a low statistical significance but it
can be seen also in other works that show the bar fraction as function of mass such as \citet{nairbfrac}, \citet{abreu12}. 
Although its low significance, such a decrease is possibly consistent with other two possible
scenarios: i) more massive disks develops a bar at earlier times with respect to their lower mass counterparts and therefore 
undergo buckling instability earlier and dismantle the bar at earlier epochs. ii) More massive disks have a different merger history
with respect of the low mass population and this may induce a different bar fraction.
In the future it would be possible to investigate these hypothesis with the advent of new cosmological simulations
at sufficient resolution.

Summarizing, in this paper I have developed an IDL-based bar finder  
that performs isophotal fitting on SDSS images and on the basis of the extracted radial ellipticity an P.A. profiles and
recognizes barred galaxies avoiding visual inspection of neither the images nor the profiles.
This procedure makes use of the tasks described in C16 and has been tested over the same sample in
order to evaluate the bar fraction in the Local and Coma supercluster and quantify th influence of barred galaxies on the average properties
of color profiles of LTGs shown in C16.

i) The procedure have extracted a fairly pure sample of barred galaxies among face-on LTGs and led to the calculation of a bar fraction of $\sim36\%$
consistent with other literature results \citep{jo04,marinova07,nairbfrac}. 
ii) The bar fraction shows a strong mass dependency 
obtained also by previous works in the local volume and at higher redshifts \citep{marinova07,nairbfrac,skibba12,abreu12}.
iii) The bars that we extracted typically occupy the central $\sim 30-40 \%$ of the host galaxy  and is typically strong ($\sim90 \%$ of times), 
consistent with the proportions observed by \citet{marinova07}.
iv) I constructed color average profiles of barred galaxies in different bins of mass and compared it to the template profiles of C16.

ii) and iii) imply that bars likely have a strong impact on the average color profiles created by C16
who observed in the template profiles of LTGs the growth of a red and dead component in an intermediate zone
inside 0.3 Petrosian radii whose importance increases with mass. 
From iv) I was able to assess that bars are redder structures with respect of their disks and can indeed reproduce the
upturn toward red of the templates profiles of C16.
Moreover this further links the presence of a bar to a decrease of the SFR in a disk galaxies as proposed by \citet{cheung13,pg15}.

\begin{acknowledgements}
I thank the Referee for the constructive criticism that helped improving
the manuscript. I acknowledge Dr. M{\'e}ndez-Abreu that kindly shared his classification 
for the comparison to this work and
Dr. Massimo Dotti and Prof. Giuseppe Gavazzi for useful discussion. 
This research has made use of the GOLDmine database (Gavazzi et
al. 2003, 2014b) and of the NASA/IPAC Extragalactic Database (NED) which is
operated by the Jet Propulsion Laboratory, California Institute of Technology,
under contract with the National Aeronautics and Space Administration.  We
wish to thank an unknown referee whose criticism helped improving the
manuscript.  Funding for the Sloan Digital Sky Survey (SDSS) and SDSS-II has
been provided by the Alfred P. Sloan Foundation, the Participating
Institutions, the National Science Foundation, the U.S. Department of Energy,
the National Aeronautics and Space Administration, the Japanese
Monbukagakusho, and the Max Planck Society, and the Higher Education Funding
Council for England.  The SDSS Web site is \emph{http://www.sdss.org/}.  The
SDSS is managed by the Astrophysical Research Consortium (ARC) for the
Participating Institutions.  The Participating Institutions are the American
Museum of Natural History, Astrophysical Institute Potsdam, University of
Basel, University of Cambridge, Case Western Reserve University, The
University of Chicago, Drexel University, Fermilab, the Institute for Advanced
Study, the Japan Participation Group, The Johns Hopkins University, the Joint
Institute for Nuclear Astrophysics, the Kavli Institute for Particle
Astrophysics and Cosmology, the Korean Scientist Group, the Chinese Academy of
Sciences (LAMOST), Los Alamos National Laboratory, the Max-Planck-Institute
for Astronomy (MPIA), the Max-Planck-Institute for Astrophysics (MPA), New
Mexico State University, Ohio State University, University of Pittsburgh,
University of Portsmouth, Princeton University, the United States Naval
Observatory, and the University of Washington.\\
\end{acknowledgements}

\end{document}